\documentclass[letterpaper, 10 pt, conference]{ieeeconf}  

\IEEEoverridecommandlockouts                              
\usepackage{amsmath} 
\usepackage{amssymb}  
\usepackage{algorithm} 
\usepackage{algpseudocode} 
\usepackage{tabularx,booktabs,mathtools,multirow,cite}

\title{\LARGE \bf
RL-Controller: a reinforcement learning framework \\for active structural control
}

\author{Soheila Sadeghi Eshkevari, Soheil Sadeghi Eshkevari, Debarshi Sen, and Shamim N. Pakzad
\thanks{This work has been submitted to the IEEE for possible publication. Copyright may be transferred without notice, after which this version may no longer be accessible.}
\thanks{SSE, DS, and SNP are with Department of Civil and Environmental Engineering, Rossin College of Engineering,
        Lehigh University, 27 Memorial Dr W, Bethlehem, PA 18015
        {\tt\small \{sos318,des620,snp208\}@lehigh.edu}}%
\thanks{SSE (second author) is with the Senseable City Lab, MIT, MIT 9-216, 77 Massachusetts Avenue, Cambridge, MA 02139 
        {\tt\small ssadeghi@mit.edu}}%
}
\usepackage{graphicx}

\begin{document}

\maketitle
\thispagestyle{empty}
\pagestyle{empty}

\begin{abstract}
To maintain structural integrity and functionality during the designed life cycle of a structure, engineers are expected to accommodate for natural hazards as well as operational load levels. Active control systems are an efficient solution for structural response control when a structure is subjected to unexpected extreme loads. However, development of these systems through traditional means is limited by their model dependent nature. Recent advancements in adaptive learning methods, in particular, reinforcement learning (RL), for real-time decision making problems, along with rapid growth in high-performance computational resources, help structural engineers to transform the classic model-based active control problem to a purely data-driven one.  In this paper, we present a novel RL-based approach for designing active controllers by introducing \emph{RL-Controller}, a flexible and scalable simulation environment. The RL-Controller includes attributes and functionalities that are defined to model active structural control mechanisms in detail. We show that the proposed framework is easily trainable for a five story benchmark building with 65\% reductions on average in inter story drifts (ISD) when subjected to strong ground motions. In a comparative study with LQG active control method, we demonstrate that the proposed model-free algorithm learns more optimal actuator forcing strategies that yield higher performance, e.g., 25\% more ISD reductions on average with respect to LQG, without using prior information about the mechanical properties of the system. 

\end{abstract}

\section{INTRODUCTION} 
Over the past few decades, structural systems have been designed and built in accordance to the well-established and experimentally-validated design codes. The resulting buildings are resistant to the most probable operational and extreme demands. However, the efficacy of the structural design has been mostly evaluated in terms of their structural resistance and human life cost prevention. In recent years, the attention has shifted towards life-cycle cost, environmental sustainability, and resiliency of the built environment. In the context of enhanced resiliency, structural control has been a long-standing solution for vibration mitigation in buildings and bridges. A controlled building has multiple orders of magnitude lower cost of rehabilitation and recovery as a consequence of its restricted demands \cite{spencer2003state}. In this aspect, the level of control determines the benefit. Traditionally, structures are equipped with passive mechanical devices that modify strength and stiffness which are important factors in counteracting demands. Still, passive control systems have limited potential in controlling motions and generally undergo residual deformations. Active control systems have been popular in mechanical and aerospace engineering for vibration control and widely used in vehicles and aircrafts \cite{zhao2018review,block1998applied,elliott2008review,yao2002mr}. In structural engineering, active control has been proposed and practically experimented; yet, due to the sensitivity and dimensional complexity of the problem, it is cautiously progressing \cite{spencer2003state}. In addition, there have been concerns regarding the capability of the active control methods in performing robustly in noisy and uncertain environments \cite{spencer1994reliability,zhang2014robust}. 

Despite the challenges and criticism, active control systems have the potential to overcome the limitations of its passive counterparts. In the active control scenario, the controllers (e.g., actuators in a building) react in accordance to a computer program that processes the real-time feed from sensory devices. In other words, in active control systems, the agents are integrated in a feedback loop and apply reactions based on the momentary demands. The majority of existing active control solutions are closed-form derivations that are based on optimal control theory which require a full description of the system subject to control \cite{balas1979direct,yang2017active}. For instance, optimal control algorithms such as linear quadratic Gaussian (LQG) control require the characteristic matrices of the state-space model of the system. This is a limiting constraint for existing structures as well as for evolving systems (e.g., due to deterioration) since the mechanical properties are not fully known. In addition, such solutions are prone to instability when the real-world structure deviates from the modeling assumptions. Therefore, an adaptive data-driven solution is highly desired.

In recent years and by proliferation of high-performance computing, data-driven methods have grown exponentially and resulted in unprecedented achievements in engineering. In particular, reinforcement learning (RL) framework has been exploited for adaptive control in robotic systems \cite{kober2013reinforcement,vecerik2017leveraging}. In this framework, the control problem is formulated as a sequential decision-making modeled with a Markov decision process (MDP). For low-complexity and discrete systems, an optimal policy can be found by constituting a Q-table for the MDP using the Bellman equation and taking the most valued action in each state (i.e., dynamic programming). The dynamic programming approach, however, is not widely applicable for real-world tasks due to the unknown nature of the transition probabilities and close-form of the reward functions. Furthermore, in complex and continuous state-action spaces, the closed-form approach would be intractable. In recent years, due to the major breakthroughs in technology and in particular, computational processing, RL has been widely utilized as a data-driven approach for estimating the optimal policy in MDPs. 

\begin{figure}[!h]
    \centering
    \includegraphics[width=80mm]{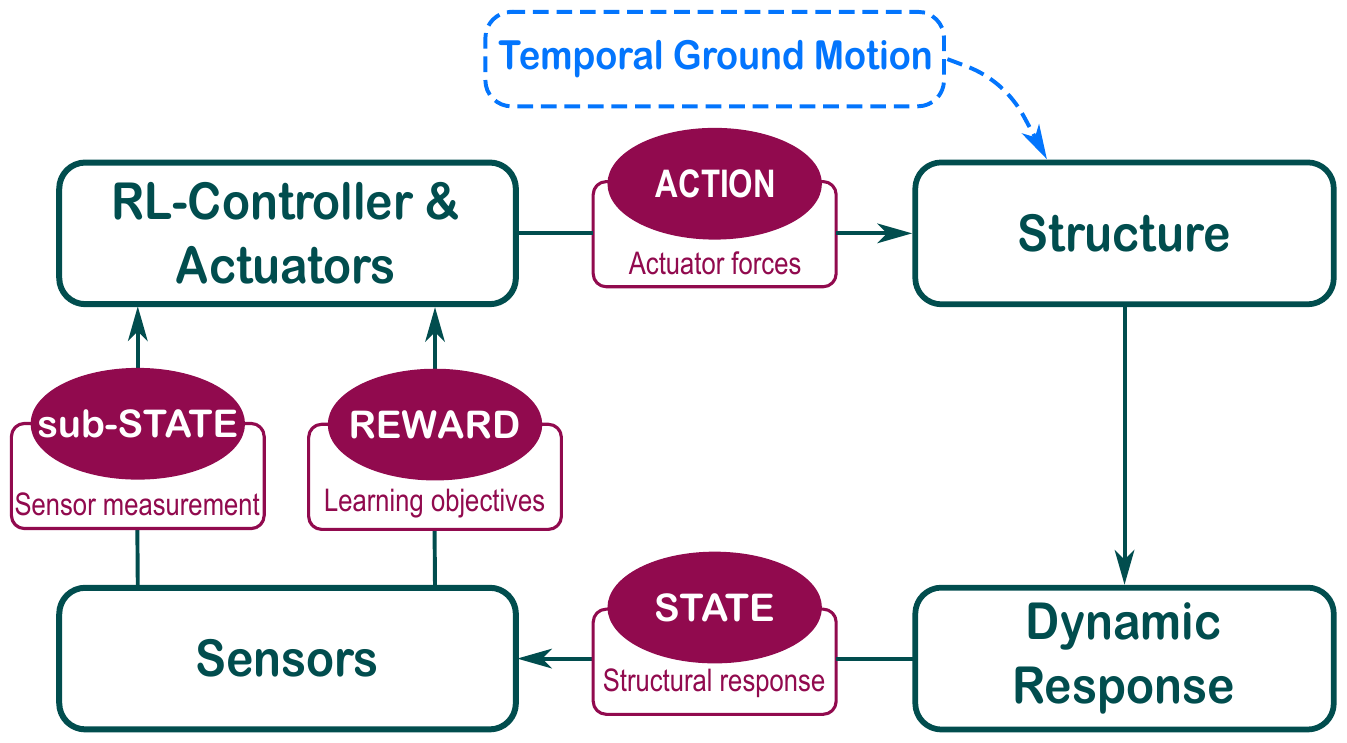}
    \caption{RL flowchart: a structural system subjected to an earthquake and an initial control force responds dynamically. Sensors capture a subset of the response which constitutes the RL state. The reward is calculated for the resulting system based on pre-determined performance objectives. Subsequently, the RL-Controller learns by exploring the action space and experimenting different trajectories.}
    \label{fig:RLFlowchart}
\end{figure}

In a RL framework, there are some principle components that define the problem, regardless of the optimization approach: (a) agent, (b) state, (c) reward, and (d) action. In deep RL, a neural network performs as the agent in order to extract useful information for the decision making process (i.e., a value function approximator). For each decision-making step, the agent processes the information regarding the existing state of the system and selects the corresponding action that is likely to maximize the cumulative reward over the course of the decision-making time frame. Using neural networks, the complexity issue in high-dimensional and continuous problems is addressed by function approximation. In our particular application, we define the RL principle components as shown in Figure \ref{fig:RLFlowchart}. In an active structural control problem, the exogenous input is defined by the ground motion (or wind loads in case of wind response control). The building structure is the environment that reacts to the input and responds dynamically. A subspace of the full state space of the building response is monitored by the sensor network (here, a network of accelerometers) and will be used as the state vector in the RL framework. In addition, using the sensory measurements, a computer system calculates the immediate reward for each discrete time step. The reward function incentivizes smaller deformation, vibrations, and actuator forces. Given the sub-state vector, an active controller that is trained with RL, predicts the optimal action: the associated actuator forces for a given time. The actuator forces and the ground motion acceleration in the new time step are input to the structure for the new cycle. 


\section{LITERATURE REVIEW} 
\subsection{Active Control in structural engineering}
Active control in the context of civil infrastructure has garnered significant interest for many decades now \cite{control_housner,control_bfs_sn}. Although passive response control devices such as dampers \cite{visco_mazza} and base isolation \cite{baseiso_buckle} devices have been used extensively for retrofitting civil infrastructure, they are limited by their design capabilities. Active control, on the other hand, allows for flexibility with regards to unknown external loads that a structure of interest is subjected to \cite{active_review1}.

Active control devices generate control forces through external power supplies that help regulate structural response. Typical control devices deployed on structures include active tendon systems \cite{at_bfs}, and active mass dampers \cite{amd_canton} to name a few. The control force generated is computed based on measurements of the external loads and/or structural response \cite{ac_review_soong}. The physical interpretation of deploying a control device is the modification of system properties and the external load, such that the structural response is minimized. Hence, the goal of a control algorithm is to estimate these modifications. 

Optimal control algorithms such as linear quadratic regulators (LQR) and linear quadratic Gaussian (LQG) controllers have been successfully applied in the context of civil engineering \cite{lqr_eg1}. Other algorithms that have been studied over the years include $H_2$, $H_{\infty}$, pole placement and sliding mode control (SMC) \cite{ac_review_casciati}. All these approaches minimize a cost function parameterized by a set of structural responses and estimated control forces acting on a structure. Additional parameters of these cost functions include user-defined weights that govern the minimization strategy. These algorithms were further enhanced by the inclusion of signal processing frameworks such as wavelet analysis \cite{ac_wavelet}, and soft computing tools like fuzzy logic, genetic algorithms and neural networks to deal with cases involving time-varying behavior, nonlinearities and uncertainties \cite{ac_fuzzy,ac_nn}. While these algorithms do not guarantee optimality, they help develop more versatile and robust control systems. 
Although soft computing tools such as neural networks functioning as model-free alternatives to traditional control have demonstrated their efficacy, they were limited to low dimensional problems owing to the curse of dimensionality and limited computational resources. As discussed earlier, recent advancements in high performance computations have revolutionized the applications of deep neural networks to solve complex problems that were deemed computationally intractable earlier. In this context, RL with embedded deep neural networks has emerged as an efficient means to achieve vibration control in engineering systems.    

\subsection{Application of RL in control}
Reinforcement learning (RL) has been a popular algorithm for solving sequential decision making problems. RL has been applied to control engineering in the context of mechanical engineering systems such as vehicles \cite{rl_vehicle}, shape control in tensegrity structures \cite{rl_tensegrity}, as well as in robotics \cite{rl_review_robotics}. However, RL in its traditional form was limited to low dimensional problems. In the past decade, equipped with better computational resources and big data, RL has been applied to far more complicated control problems such as active flow control in computational fluid dynamics \cite{rl_cfd}, bluff body flow control \cite{rl_bluffbody}, learning robotic locomotion \cite{rl_roboticloco}, vision-based robotic manipulation \cite{pmlr-v87-kalashnikov18a}, and mapless robot navigation \cite{8202134}, to name a few. 

In the context of active structural control there have been limited studies on application of RL. For example, RL was used to tune a fuzzy logic control-based active mass damper recently \cite{rl_fuzzy}. However, with the advent of better computational resources and the suitability of the structural control problem to be formulated as a MDP problem, RL has the potential to be a powerful tool for active structural control. In this paper, we propose RL-Controller, a framework that generalizes the application of RL for structural control problems. In the following sections we describe the proposed framework and demonstrate its efficacy through a numerical example. 

\section{METHODOLOGY} 
In this paper, we first propose a flexible and scalable Python class\footnote{According to Python documentation, classes provide a means of bundling data and functionality together.} to utilize RL in structural control problems adaptable to buildings. We define this interface as a standard \emph{Gym} environment to conform with common RL applications while seeking to solve a real-life concern of structural engineers. Gym library is a well-known user-friendly interface defined for RL applications such as CartPole, MountainCar and Pendulum \cite{brockman2016openai}.

\subsection{RL-Controller class}
\begin{figure}[!h]
    \centering
    \includegraphics[width=85mm]{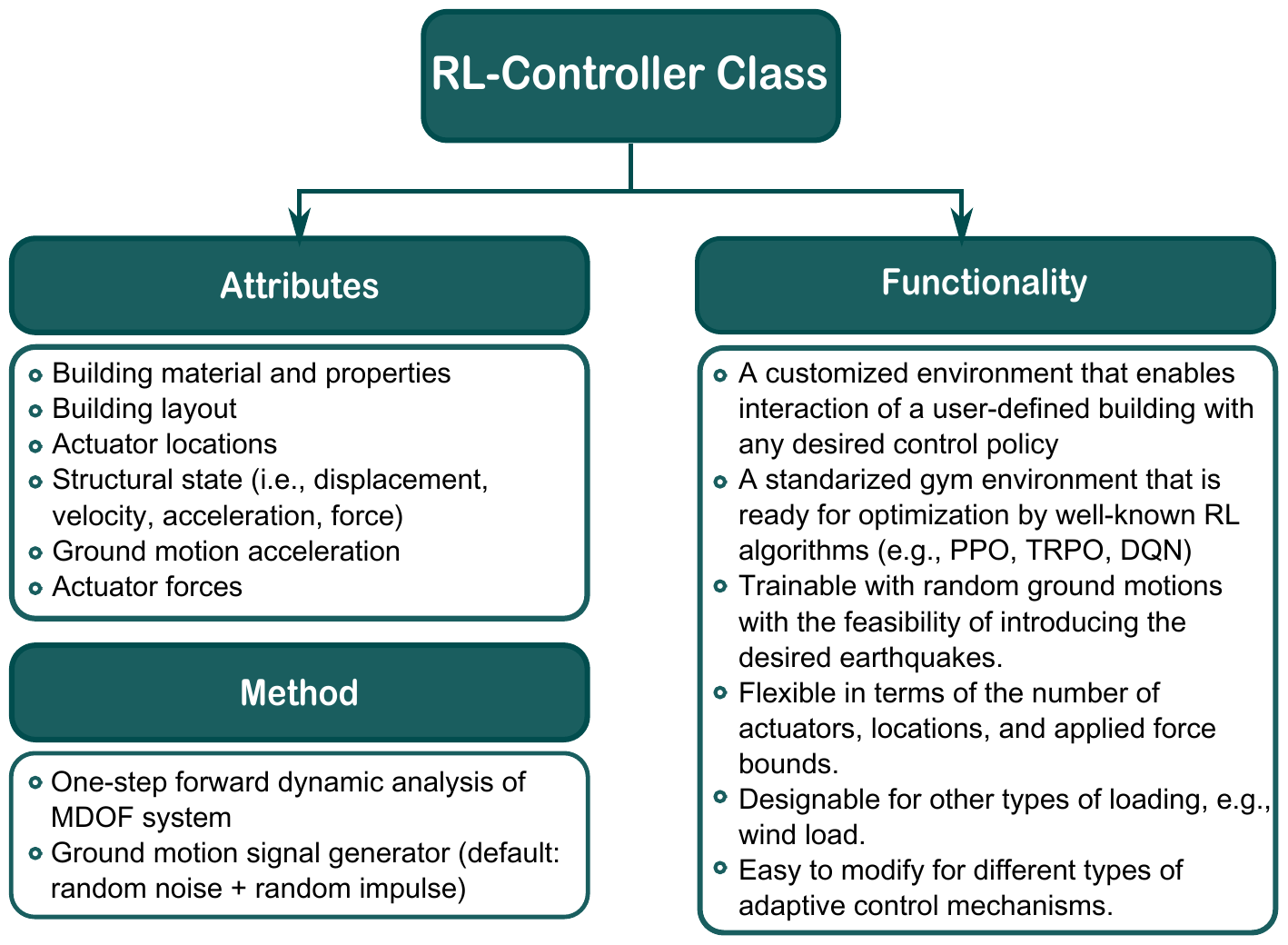}
    \caption{RL-Controller class description: RL-Controller is a designed interface in Gym environment to address structural control problem with RL.}
    \label{fig:ClassFlowchart}
\end{figure}

In this section, we discuss the attributes, methods and designed functionalities of the proposed class as shown in Figure \ref{fig:ClassFlowchart}:

\textbf{Attributes:} The RL-Controller class takes geometric and material properties of the structure of interest, as well as the planned actuator layout that imparts the control force as an input. To maintain a model-free controller, the user must measure structural responses such as story accelerations in this platform or has to construct a trustworthy model for data generation purposes. The state vector in RL setup can be defined based on the sensory devices that are available. By default, the RL state includes acceleration responses in different DOFs and at multiple most recent time steps capturing the inherent dynamics of the system. Consequently, the length of the structural response history governs the performance of the learning-based agent in estimating the control force. Hence, we treat the number of time steps as a hyperparameter. The class estimates control forces in real time for the structure based on the ground motion acceleration it is subjected to. These control forces act on the structure simultaneously with the external loads.

\textbf{Method:} To model the dynamic behaviour of a building for data generation step, the  RL-Controller uses numerical simulation of dynamic response with Newmark-$\beta$ Method for multi degree of freedom (MDOF) systems \cite{chopra}. For the training process, the agent randomly explores the environment by applying various possible control forces that an actuator can impart on the structure. This ensures that the controller is prepared for any unforeseen external load that the structure may encounter in the future. To enhance robustness, we define a ground motion generator to apply random noise added to random impulses during the training process. This loading strategy is found to be simple, minimal, and effective when the trained structure is subjected to real earthquake records.

\textbf{Functionality:} The proposed class RL-Controller is a customized Gym environment with flexible user-defined parameters. The defined class can be used for learning the optimal control policy using many novel optimization methods to meet the user's needs. The training process is designed to be efficient while it is capable of handling any customized structural demands, i.e., desired ground motions. Same as any Gym environment, the RL-Controller class is flexible to changes in attributes or optimization parameters to match the problem and action levels the best. The algorithm can assign different sub-space of the building full response as the RL state vector (e.g., acceleration, velocity, displacement, force, or any combination or subset of these). It is also possible to simply modify the environment for different loading regimes such as wind load. Furthermore, the class enables various control mechanisms to be introduced in the system (e.g., actuator force, tendon mechanism, damping tuning systems, etc.). 

\subsection{RL-Controller's training and optimization}

The RL framework introduced in this study enables off-policy and on-policy learning. For the on-policy setting, one desired functionality is an integrated structural dynamic simulation for data generation purposes based on user-defined structural properties. Note that the simulator does not contribute to the learning process and merely acts as a problem simulation environment, as it is common in similar RL settings. In this section we draw a clear outline of how the RL attributes are defined in the Gym environment. The \emph{state} (distinct from state as defined in control theory wherein it refers to the state space variables of a dynamic system) in this simulation (Equation \ref{eq:state}) is a function ($f^s$) of the acceleration of the instrumented levels, ground acceleration at last time step, and the most recent applied actuator forces. Choosing acceleration for inclusion in the state was mainly based on accessibility of accelerometers and their installation convenience compared to other response signals. This state function is found sufficiently informative to guide the RL agent for controlling the system. The objective function for the optimization problem of the RL class is captured by a reward function as shown in Equation \ref{eq:reward}, which we define as a function ($f^r$) of displacements, base shear, and applied force. Actions determined from the RL-Controller's final policy, are the applied control forces on structure from the deployed actuators.

\begin{equation}\label{eq:state}
    \textbf{s} = f^s(\ddot{\textbf{x}},\ddot{\textbf{x}}^g,\textbf{a})
\end{equation}

where $\textbf{s}$ represents the state, $\ddot{\textbf{x}}$ is the acceleration vector, $\ddot{\textbf{x}}^g$ is the ground acceleration and $\textbf{a}$ is the action taken by the policy.
\begin{equation}\label{eq:reward}
    \textbf{r} = f^r(\textbf{x},\textbf{m}\ddot{\textbf{x}},\textbf{a})
\end{equation}

where $\textbf{r}$ shows the reward value, $\textbf{x}$ is the story displacement vector, $\textbf{m}$ is the mass matrix for base shear calculation. The arguments of reward function require minimization for efficient response control. Therefore, a negative multiplier is applied in the reward function for the optimization process (e.g., minimizing displacement maximizes the reward). It should be noted that we scale each variable in the reward function prior to use in the reward function to promote numerical stability.

To determine the best optimization result in terms of speed and optimization values, we employed several optimization methods. The best performing methods for this problem are proximal policy optimization (PPO) and soft-actor-critic (SAC). PPO is a policy gradient technique based on trust region method which uses several epochs of sampled data while utilizing the policy \cite{schulman2017proximal}. SAC is an off-policy optimization method which seeks to maximize entropy, for broader exploration, and the expected reward \cite{haarnoja2018soft}. These two methods work effectively in this control problem and both reach a reasonable optimum in each training event. PPO reaches a more stable optimum in most training episodes, although, because of more extensive action space exploration in SAC for this high dimensional problem, it holds a better computational efficiency and the optimum was achieved considerably faster. Accordingly, all the discussions in Section \ref{sec:results} are based on SAC as the optimization method for finding the optimal policy. The architecture of the neural network that behaves as the agent is a multilayer perceptron network with three layers, each of size 128. Note that being a standard Gym environment, RL-Controller provides this flexibility to experiment with various state-of-the-art optimization algorithms and network architectures with minimum effort.

\begin{algorithm}
	\caption{RL-Controller training process}
	\label{algorithm:RL}
	\begin{algorithmic}[1]
	\State Initialization step for $f^s$, $f^r$
	    \State $s_{0}=\{\ddot{x}_{0},\ddot{x}^g_0,a_{0}\}$
		\For {each iteration}
			\For {$t=0,\Delta t,2\Delta t,\ldots,T$}
			    \State $a_{t} = \pi_{\theta}(s_{t})$
			    \State $x_{t+1},\dot{x}_{t+1},\ddot{x}_{t+1}=f^s(x_{t},\dot{x}_{t},\ddot{x}_{t},\ddot{x}^g_{t},a_{t})$
			    \State $s_{t+1}=\{\ddot{x}_{t+1},\ddot{x}^g_{t+1},a_{t+1}\}$
			    \State $r(s_{t},a_{t})=f^r(x_{t},m.\ddot{x}_{t},a_{t})$
    		    \State $\pi_{\theta}\leftarrow{SAC(\pi_{\theta},\{s_t,r(s_t,a_t),s_{t+1}\})}$.
			\EndFor

		\EndFor
	\end{algorithmic} 
\end{algorithm}

To train the RL-Controller, as shown in Algorithm \ref{algorithm:RL}, we first initialize the defined state and reward functions in Equation \ref{eq:state} and \ref{eq:reward} to start the learning iterations. Throughout the training process, the action or actuator forces are estimated based on the last updated policy (random at first). The state and action along with the given ground motion are inputs to update the state using the function $f^s$ and obtain reward of this action from the function $f^r$. In the end, the policy is updated using SAC algorithm. By repeating this training process for sufficient numbers of iterations, the optimizer is able to find a reliable optimal strategy. The subscripts in the algorithm shows the time step, $\pi_\theta$ is the policy, $\dot{x}$ is the story velocity and $SAC$ shows the optimization step taken for each training.


\section{RESULTS}\label{sec:results} 
\subsection{Benchmark five-story building model}
For evaluation of the proposed framework, we use a five-story shear benchmark building that is modeled as a linear MDOF system \cite{park2017updating}. Table \ref{tab:BuildingProperties} shows the structural properties of each degree of freedom. To model structural damping as Rayleigh damping, the damping ratios of the first and fifth modes are set to 1\% and 5\%, respectively. We assume that three actuators are located at stories one,three and five.

\begin{table}[t]
    \centering
    \caption{Five-story building model properties}
    \begin{tabular}{lcc}
    \toprule
    Story & Mass ($\times10^3$kg) & Stiffness ($\times10^6$N/m)\\
    \midrule
    One & 25 & 5\\
    Two & 20 & 4\\
    Three & 20 & 4\\
    Four & 18 & 3\\
    Five & 15 & 3\\
    \bottomrule
    \end{tabular}
    \label{tab:BuildingProperties}
\end{table}

\begin{figure}[!h]
    \centering
    \includegraphics[height=30mm]{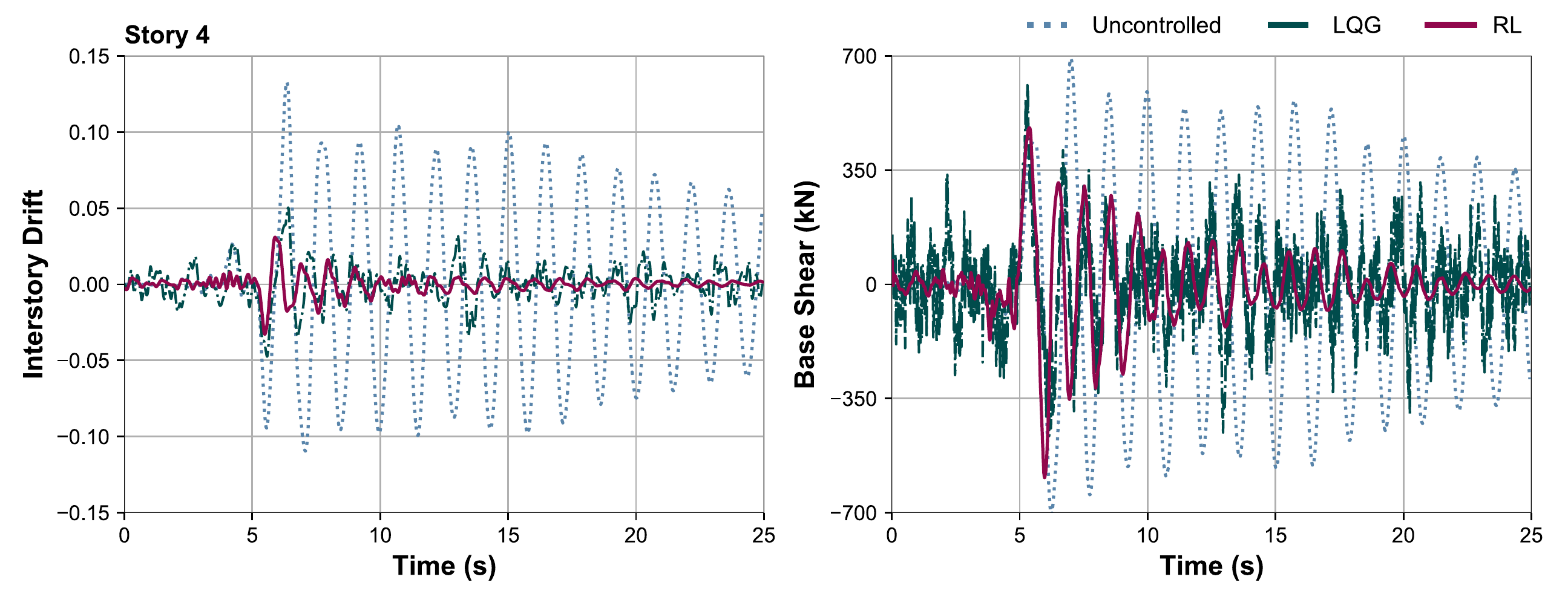}
    \caption{Comparison for response to Northridge (1994) earthquake.}
    \label{fig:Response Comparison}
\end{figure}

We train the RL-Controller class for this benchmark building for 1.5 million iterations to reach optimal policy with SAC. We observe that using five recent acceleration time history records for RL-Controller's action evaluation led to the most efficient performance. To increase long lasting action impact we set the discount factor ($\gamma$ in RL) for weighing temporally distant rewards to 0.999. Figure \ref{fig:Response Comparison} compares the inter-story drift (ISD) at story four and base shear among uncontrolled system, controlled with LQG, and the RL-Controller when the structure is subjected to the Northridge (1994) earthquake, recorded at the West Pico Canyon Road station. For a fair comparison, the LQG controller is designed such that the corresponding cost function is similar to the one used by RL-Controller. 

The ISD of fourth story shows that the RL-Controller is able to limit the deformation amplitudes substantially compared to the uncontrolled structure at stories without any installed actuators. Furthermore, the RL-Controller outperforms LQG for this earthquake in terms of drift. Figure \ref{fig:Response Comparison} shows base shear for the aforementioned three cases. We observe that in terms of base shear the RL-Controller's performance is considerably superior. Base shear is a measure of the cumulative acceleration/inertial forces acting on all stories. Acceleration and displacement in a controlled structure have a trade-off, because a high control force in favor of low displacement increases the accelerations of a structure. Yet, the plot shows the RL-Controller is able to limit ISD while maintaining a low overall acceleration and hence, base shear.  

It is common in RL that the trained agent learns efficient and smart strategies. In this problem, the RL-Controller learns to absorb the shocks (impulses) momentarily and also control both displacement and base shear simultaneously, significantly outperforming LQG. We believe that the improved performance is due to the inclusion of impulse loads along with white noise as the base excitation during the training process. This flexibility allows for a better control system design compared to LQG where external loads are assumed to be a Gaussian white noise. 


\subsection{Comparative study}
For a comprehensive assessment, we subject the benchmark structure to multiple historical earthquakes. In order to quantify performance, we define four metrics $J_1$ to $J_4$ as shown in Equations \ref{eq:J1} and \ref{eq:J3}. In the following equations, $C$ and $UC$ represent controlled and uncontrolled measurements, respectively.

\begin{equation}\label{eq:J1}
    J_{1} = \dfrac{\max |\delta_{C}|}{\max |\delta_{UC}|}, \; J_{2} = \dfrac{\max |\ddot{x}_{C}|}{\max |\ddot{x}_{UC}|}
\end{equation}

where, $\delta$ is the ISD, and $\ddot{x}$ is the story acceleration. $J_{1}$ is a ratio between the maximum ISD value at each story for each earthquake to compare controlled and uncontrolled drifts. $J_{2}$ captures the same ratio as $J_{1}$, except that it studies the accelerations instead of ISDs.

    




\begin{equation}\label{eq:J3}
    J_{3} = \dfrac{E_{u}}{E_{V_{b,UC}}}, \; J_{4} = \dfrac{\max |V_{C}|}{\max |V_{UC}|}
\end{equation}

where, $E_{u}$ is the applied control force signal energy calculated as Equation \ref{eq:energy} shows, and $E_{V_{b,UC}}$ is the energy of uncontrolled base shear. We used $J_{3}$ to compare the control force signal energy imparted to the structure normalized by the uncontrolled base shear signal energy. Also, $V$ stands for the story shear and $J_{4}$ compares the maximum controlled and uncontrolled story shears.

\begin{equation}\label{eq:energy}
    E_{x} = \int_{0}^{T} |x(t)|^2 \,dt
\end{equation}

\begin{figure}[!h]
    \centering
    \includegraphics[height=60mm]{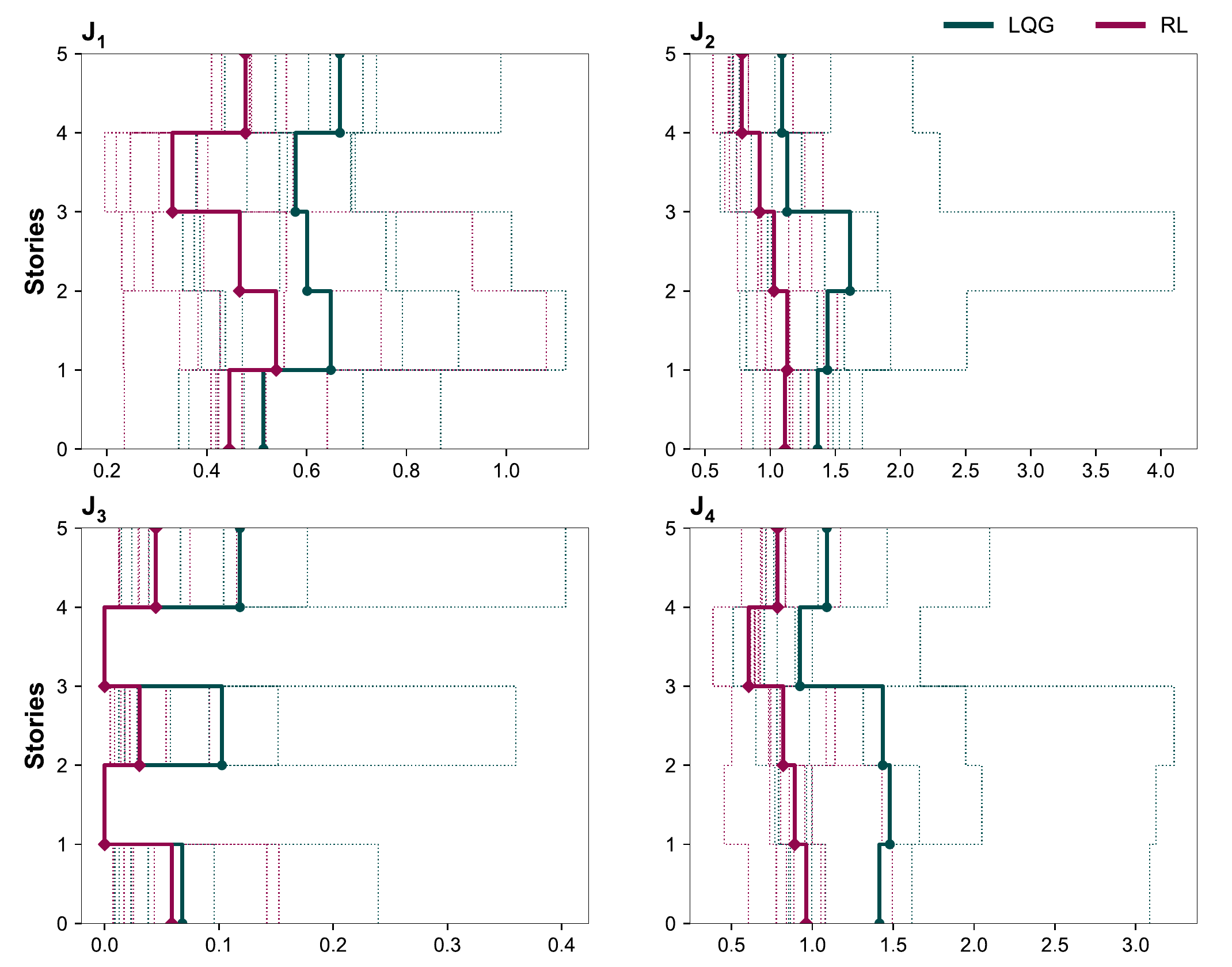}
    \caption{Metrics comparison for the five-story benchmark building. The bold lines show the mean value of the metrics averaged over seven significant earthquake records.}
    \label{fig:Metrics Comparison}
\end{figure}

Figure \ref{fig:Metrics Comparison} shows the four metrics for each story subjected to seven records of strong ground motions. The chosen earthquakes are Loma Prieta (1989), Imperial Valley (1979), Coalinga (1983), Kobe (1995), Chi Chi (1999) and Northridge (1994, Sylmar and W Pico Canyon Road stations). We selected these set of ground motions based on their variability in amplitude, frequency content, and duration to capture ground motion parameter uncertainties. The bold lines in the figures show the average metrics over all the earthquakes for both LQG and RL. In terms of structural response minimization, we observe a strictly better performance for the RL-Controller as $J_{1}$ and $J_{2}$ are on average consistently lower than LQG. From $J_{3}$, which quantifies the signal energy content of the applied control force, we observe lower control forces for the RL-Controller. The ratio of maximum controlled to uncontrolled story shear, $J_{4}$, shows a major advantage of the RL-Controller over LQG. Considering that the objective function in both methods are modeled similarly, the learning process with RL is notably advantageous on average for the set of ground motions considered.

\section{CONCLUSION}
In this paper we present RL-Controller, which is a flexible tool for data-driven and adaptive structural control applicable to buildings. We discuss the essential attributes, applied methods and significant functionalities of the proposed controller. This framework is a model-free solution for the active control problem, thus substantially expanding its applicability. In a numerical case study on a five-story benchmark building, we show that the proposed framework is successful in minimizing structural response compared to a classical active control method based on optimal control theory. On average we observe a 25\% and 26\% enhancement in performance when comparing RL-Controller to LQG for inter story drift (ISD) and accelerations, respectively. In the future we plan to study the effect of building geometry, uncertainty propagation, and different control mechanisms such as tendon and mass damper systems, on the performance of  RL-Controller.




\bibliographystyle{ieeebib}
\bibliography{references}

\end{document}